\documentclass[twocolumn]{aastex631}
\usepackage{appendix}
\usepackage{natbib}
\usepackage{subfigure}
\usepackage{booktabs}
\usepackage{xspace}
\usepackage[encapsulated]{CJK}

\usepackage{amsmath} 

\newcommand{\oiii}{[\ion{O}{3}]\xspace}

\newcommand{\msun}{$M_{\odot}$}

\newcommand{\mbh}{$M_{\rm BH}$}

\newcommand{\hide}[1]{}

\newcommand{\jw}{\emph{JWST}}
    
\usepackage[encapsulated]{CJK}

\begin{document}
   
\title{What you see is what you get: empirically measured bolometric luminosities of Little Red Dots}

\author[0000-0002-5612-3427]{Jenny E. Greene}
\affiliation{Department of Astrophysical Sciences, Princeton University, 4 Ivy Lane, Princeton, NJ 08544, USA}

\author[0000-0003-4075-7393]{David J. Setton}
\thanks{Brinson Prize Fellow}
\affiliation{Department of Astrophysical Sciences, Princeton University, 4 Ivy Lane, Princeton, NJ 08544, USA}

\author[0000-0001-6278-032X]{Lukas J. Furtak}
\affiliation{Department of Physics, Ben-Gurion University of the Negev, P.O. Box 653, Be'er-Sheva 84105, Israel}

\author[0000-0003-3729-1684]{Rohan P. Naidu}
\affiliation{MIT Kavli Institute for Astrophysics and Space Research, 70 Vassar Street, Cambridge, MA 02139, USA}

\author[0000-0002-3216-1322]{Marta Volonteri}
\affiliation{Institut d'Astrophysique de Paris, Sorbonne Universit\'{e}, CNRS, UMR 7095, 98 bis bd Arago, 75014 Paris, France}

\author[0000-0001-8460-1564]{Pratika Dayal}
\affiliation{Kapteyn Astronomical Institute, University of Groningen, P.O. Box 800, 9700 AV Groningen, The Netherlands}
\affiliation{Canadian Institute for Theoretical Astrophysics, 60 St George St, University of Toronto, Toronto, ON M5S 3H8, Canada}

\author[0000-0002-2057-5376]{Ivo Labbe}
\affiliation{Centre for Astrophysics and Supercomputing, Swinburne University of Technology, Melbourne, VIC 3122, Australia}

\author[0000-0002-8282-9888]{Pieter van Dokkum}
\affiliation{Astronomy Department, Yale University, 52 Hillhouse Ave,New Haven, CT 06511, USA}

\author[0000-0001-5063-8254]{Rachel Bezanson}
\affiliation{Department of Physics and Astronomy and PITT PACC, University of Pittsburgh, Pittsburgh, PA 15260, USA}

\author[0000-0003-2680-005X]{Gabriel Brammer}
\affiliation{Cosmic Dawn Center (DAWN), Niels Bohr Institute, University of Copenhagen, Jagtvej 128, K{\o}benhavn N, DK-2200, Denmark}

\author[0000-0002-7031-2865]{Sam E. Cutler}
\affiliation{Department of Astronomy, University of Massachusetts, Amherst, MA 01003, USA}

\author[0000-0002-3254-9044]{Karl Glazebrook}\affiliation{Centre for Astrophysics and Supercomputing, Swinburne University of Technology, PO Box 218, Hawthorn, VIC 3122, Australia}

\author[0000-0002-2380-9801]{Anna~de~Graaff}
\affiliation{Max-Planck-Institut f\"ur Astronomie, K\"onigstuhl 17, D-69117 Heidelberg, Germany}

\author[0000-0002-3301-3321]{Michaela Hirschmann}
\affiliation{Institute of Physics, GalSpec laboratory, EPFL, Observatory of Sauverny, Chemin Pegasi 51, 1290 Versoix, Switzerland}

\author[0000-0002-4684-9005]{Raphael E. Hviding}
\affiliation{Max-Planck-Institut f\"ur Astronomie, K\"onigstuhl 17, D-69117 Heidelberg, Germany}

\author[0000-0002-5588-9156]{Vasily Kokorev}
\affiliation{Department of Astronomy, The University of Texas at Austin, Austin, TX 78712, USA}

\author[0000-0001-6755-1315]{Joel Leja}
\affiliation{Department of Astronomy \& Astrophysics, The Pennsylvania State University, University Park, PA 16802, USA}
\affiliation{Institute for Computational \& Data Sciences, The Pennsylvania State University, University Park, PA 16802, USA}
\affiliation{Institute for Gravitation and the Cosmos, The Pennsylvania State University, University Park, PA 16802, USA}

\author[0000-0003-2488-4667]{Hanpu Liu}
\affiliation{Department of Astrophysical Sciences, Princeton University, 4 Ivy Lane, Princeton, NJ 08544, USA}

\author[0000-0002-0463-9528]{Yilun Ma (\begin{CJK*}{UTF8}{gbsn}马逸伦\ignorespacesafterend\end{CJK*})}
\affiliation{Department of Astrophysical Sciences, Princeton University, 4 Ivy Lane, Princeton, NJ 08544, USA}

\author[0000-0003-2871-127X]{Jorryt Matthee}
\affiliation{Institute of Science and Technology Austria (ISTA), Am Campus 1, Klosterneuburg, Austria}

\author[0000-0003-2804-0648 ]{Themiya Nanayakkara}
\affiliation{Centre for Astrophysics and Supercomputing, Swinburne University of Technology, PO Box 218, Hawthorn, VIC 3122, Australia}

\author[0000-0001-5851-6649]{Pascal A.\ Oesch}
\affiliation{Department of Astronomy, University of Geneva, Chemin Pegasi 51, 1290 Versoix, Switzerland}
\affiliation{Cosmic Dawn Center (DAWN), Copenhagen, Denmark}
\affiliation{Niels Bohr Institute, University of Copenhagen, Jagtvej 128, K{\o}benhavn N, DK-2200, Denmark}

\author[0000-0002-9651-5716]{Richard Pan}\affiliation{Department of Physics and Astronomy, Tufts University, 574 Boston Ave., Medford, MA 02155, USA}

\author[0000-0002-0108-4176]{Sedona H. Price}
\affiliation{Space Telescope Science Institute (STScI), 3700 San Martin Drive, Baltimore, MD 21218, USA}

\author[0000-0003-3256-5615]{Justin~S.~Spilker}
\affiliation{Department of Physics and Astronomy and George P. and Cynthia Woods Mitchell Institute for Fundamental Physics and Astronomy, Texas A\&M University, 4242 TAMU, College Station, TX 77843-4242, US}

\author[0000-0001-9269-5046]{Bingjie Wang (\begin{CJK*}{UTF8}{gbsn}王冰洁\ignorespacesafterend\end{CJK*})}
\affiliation{Department of Astronomy \& Astrophysics, The Pennsylvania State University, University Park, PA 16802, USA}
\affiliation{Institute for Computational \& Data Sciences, The Pennsylvania State University, University Park, PA 16802, USA}
\affiliation{Institute for Gravitation and the Cosmos, The Pennsylvania State University, University Park, PA 16802, USA}

\author[0000-0003-1614-196X]{John R. Weaver}
\affiliation{Department of Astronomy, University of Massachusetts, Amherst, MA 01003, USA}

\author[0000-0001-7160-3632]{Katherine E. Whitaker}
\affiliation{Department of Astronomy, University of Massachusetts, Amherst, MA 01003, USA}
\affiliation{Cosmic Dawn Center (DAWN), Denmark}

\author[0000-0003-2919-7495]{Christina C.\ Williams}
\affiliation{NSF National Optical-Infrared Astronomy Research Laboratory, 950 North Cherry Avenue, Tucson, AZ 85719, USA}

\author[0000-0002-0350-4488]{Adi Zitrin}
\affiliation{Department of Physics, Ben-Gurion University of the Negev, P.O. Box 653, Be'er-Sheva 84105, Israel}
 
\date{May 2025}

\begin{abstract}
New populations of red active galactic nuclei (known as ``Little Red Dots'') discovered by JWST exhibit remarkable spectral energy distributions. Leveraging X-ray through far-infrared observations of two of the most luminous known Little Red Dots, we directly their bolometric luminosities. We find evidence that more than half of the bolometric luminosity likely emerges in the rest-frame optical, with $L_{\rm bol}/L_{5100} = 5$, roughly half the value for ``standard'' Active Galactic Nuclei. Meanwhile, the X-ray emitting corona, UV-emitting black-body, and reprocessed mid to far-infrared emission are all considerably sub-dominant, assuming that the far-infrared luminosity is well below current measured limits. We present new bolometric corrections that dramatically lower inferred bolometric luminosities by a factor of ten compared to published values in the literature. These bolometric corrections are in accord with expectations from models in which gas absorption and reprocessing are responsible for the red rest-frame optical colors of Little Red Dots. We discuss how this lowered luminosity scale suggests a lower mass scale for the population by at least an order of magnitude (e.g., $\sim 10^5-10^7$~\msun\ black holes, and $\sim 10^8$~\msun galaxies), alleviating tensions with clustering, overmassive black holes, and the integrated black hole mass density in the Universe.
\end{abstract}


\section{Introduction}
\label{sec:intro}

\jw\ has unveiled a new population of compact high-redshift sources colloquially known as Little Red Dots \citep{Matthee:2024}. These objects have generated enormous excitement because of their high number densities \citep{Labbe:2023uncover,Kokorev:2024,Kocevski:2024} and puzzling spectral energy distributions \citep[e.g.,][]{Barro:2023,Williams:2024,Yue:2024}.

The first Little Red Dots were identified based on their compact sizes, red rest-frame optical colors, and faint blue UV continua \citep[e.g.,][]{Labbe:2023,Furtak:2023,Labbe:2023uncover,Hviding:2025}. Due to their small size and extreme redness, they were thought to be powered by dust-reddened active galactic nuclei (AGN) or massive galaxies \citep[e.g.,][]{Barro:2023,Labbe:2023,Baggen:2024}. Subsequent spectroscopy demonstrated a high fraction of broad Balmer emission lines \citep{Harikane:2023,Matthee:2024,Furtak:2023nature,Kokorev:2023,Greene:2024,Wang:2024brd,Lin:2025}, seemingly supporting the accreting black hole hypothesis. Note that compact red sources can also be powered by dusty starforming galaxies, particularly since high equivalent width (EW) H$\alpha$ or [OIII] emission lines can boost the broad-band photometry and cause the rest-frame optical color to appear red \citep[e.g.,][]{PerezGonzalez:2024,Hviding:2025}.

Little Red Dots are inferred to account for a substantial fraction of the broad-line AGN population \citep{Harikane:2023,Hviding:2025}, and their bolometric luminosities provide an important clue to their nature. At first, the redness that characterizes Little Red Dots was assumed to arise from dust in front of a typical UV-bright AGN or a star-forming galaxy. In the AGN scenario, to calculate the total bolometric luminosity one would take a bolometric correction from the literature for standard AGN \citep[e.g. based on the H$\alpha$ luminosity][]{GreeneHo:2005}, and then apply a significant reddening correction for $A_V \sim 2-5$~mag. The resulting inferred bolometric luminosities in the literature are $L_{\rm bol} \sim 10^{44}-10^{46}$~erg/s for typical objects \citep[e.g.,][]{Matthee:2023,Lin:2024aspire}, which corresponds to the Eddington limit for $\sim 10^7$~\msun\ black holes.  

However, broad-band spectral energy distributions quickly complicate the assumption of dust-reddened, but otherwise normal, AGN. The lack of X-ray emission \citep{Furtak:2023nature,Yue:2024,Ananna:2024}, at levels at least 10--100 times weaker than local accreting black holes, seems to cast doubt on the AGN explanation. The rising red continua observed in the rest-frame optical do not continue to the mid-infrared as seen in nearly all AGN \citep{Williams:2024, Wang:2024brd, Setton:2025, deGraaff:2025}. Instead, the sources flatten in $f_{\nu}$ between 0.7$\micron$ and rest-frame 3--5$\micron$, the reddest wavelengths that have been robustly detected with MIRI for individual sources to date. Thirdly, there is not evidence for significant reprocessed emission in the far-infrared, as would be expected if the red rest-frame optical color arises from attenuation by dust \citep{Akins:2024,Setton:2025,Xiao:2025,Casey:2025,Chen:2025dust}. There are recent models that explore attenuation by a dense envelope of hydrogen surrounding an embedded source \citep{Inayoshi:2024,Ji:2025,Naidu:2025,deGraaff:2025,Liu:2025lrd}, but in this paper we mostly focus on a purely empirical characterization of the emergent observed spectral energy distribution (see \S \ref{sec:models}).

Given tremendous progress in measuring the spectral energy distributions of luminous Little Red Dots, we are now in a position to directly measure the bolometric luminosities rather than infer them. Here we will not assume a dust-reddened UV-bright AGN to derive bolometric luminosities, nor will we assume standard scaling relations between H$\alpha$ and bolometric luminosity can be applied. Instead, we will take an empirical approach and measure the bolometric luminosity. We focus on two luminous sources with very complete SED information (\S \ref{sec:data}) and revisit the bolometric luminosities of Little Red Dots (\S \ref{sec:sed}). We propose new bolometric corrections (\S \ref{sec:bolometric}), investigate the remaining systematic uncertainties (\S \ref{sec:systematics}), and explore the ramifications of this downward shift in bolometric luminosity (\S \ref{sec:demographics}). We will discuss modeling a bit further in \S \ref{sec:models}, but our main focus in the bulk of the paper is to explore bolometric corrections based on the observed spectral energy distribution with no dust correction.

\begin{figure*}
\vspace{-10mm}
\centering
\includegraphics[width=0.85\textwidth]{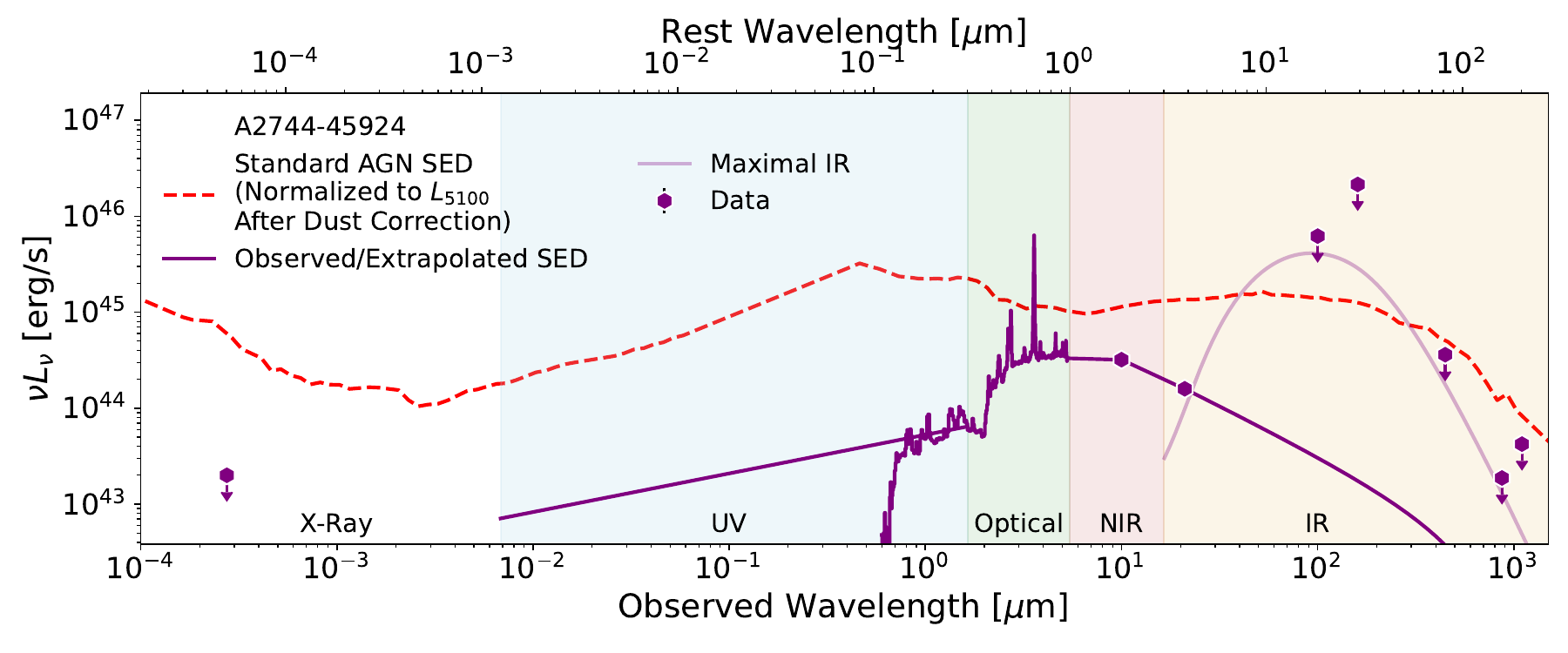}
\includegraphics[width=0.85\textwidth]{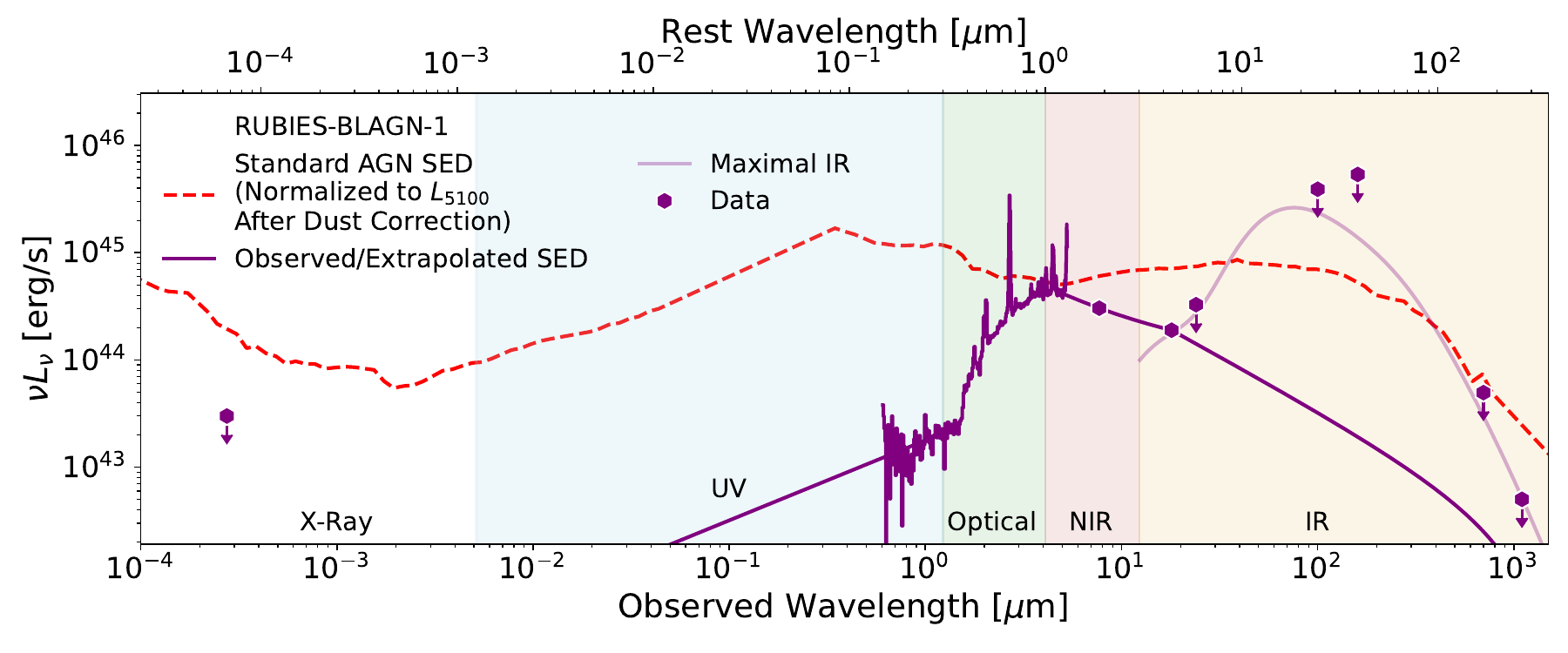} 
\caption{The panchromatic spectral energy distributions of A2744-45924 ($z=4.46$; top) and RUBIES-BLAGN-1 ($z=3.1$, bottom). These are the SEDs that we integrate to derive bolometric luminosities. The bolometric corrections are calculated in the shaded (rest-frame) regions labeled as X-ray (white), UV (blue), optical (green), NIR (red), and IR (goldenrod) following the definitions in \cite{RisalitiElvis:2004} that we adopt in Table \ref{tab:bolcor}. The NIRSpec/PRISM spectrum \cite{Labbe:2025} and \cite{Wang:2024brd} are plotted in purple, as are the MIRI detections and ALMA limits from \cite{Setton:2025} along with the Spitzer/MIPS 24 $\mu$m limit for RUBIES-BLAGN-1 from \citet[][]{Wang:2024brd}, and the X-ray upper limits. We also extrapolate the rest-UV, the rest-NIR, the minimum FIR (dark purple solid line), and the upper limit for the FIR SED used to calculate the bolometric luminosity (which we show as a light purple shaded line to distinguish it from the rest of the SED that goes into our minimum-FIR $L_{bol}$ calculation). We also show a standard AGN SED (dashed red line; accreting at $\sim$the Eddington limit) from \cite{ho2008}, scaled to the $L_{5100}$ we calculated based on an assumed $A_v=1.5$, the typical correction assumed in \cite{Greene:2024}. In prior work, this dashed red line has been assumed to be intrinsic to LRDs, and then dust reddened. The deep ALMA limits clearly rule out such an SED.}
\label{fig:MonsterSED}
\end{figure*}

\section{Samples and Data}
\label{sec:data}

\subsection{A2744-45924 and RUBIES-BLAGN-1}

A2744-45924 and RUBIES-BLAGN-1 are two of the most luminous known spectroscopically identified Little Red Dots, and form the basis of this work. A2744-45924 was identified from the photometric selection of \citet{Labbe:2023uncover}. It satisfies the spectroscopic selection of \citet{Hviding:2025}, but because it is relatively blue in the rest-frame optical, it is excluded by the selection of \citet{Kocevski:2024}. RUBIES-BLAGN-1 was initially targeted by the RUBIES program based on NIRCAM/F150$-$F444 color \citep{Degraaff:2024RUBIES}, but of the Little Red Dot selections, only that of \citet{Kocevski:2024} recovers it. RUBIES-BLAGN-1 would not be picked up by \citet{Kokorev:2024}, \citet{Barro:2024select}, or the red color cut from \citet{Greene:2024} used by \citet{Akins:2024}. 

\citet{Setton:2025} present deep MIRI and ALMA observations for two of the most optically luminous known Little Red Dots \citep[see also ][]{Akins:2024}. \citet{Labbe:2025} present an in-depth analysis of the PRISM spectrum \citep{Price:2024}, medium-band imaging \citep{Suess:2024}, and grism spectrum \citep{Naidu24} of A2744-45924 ($z=4.46$) which was selected from UNCOVER \citep{Bezanson:2022} imaging in \citet{Labbe:2023uncover}. \citet{Torralba:2025} present a detailed study of the Ly~$\alpha$ emission from A2744-45924, which is offset from the rest-optical point source and is likely to be associated with the host. That extended light is not included in our analysis. \citet{Wang:2024brd} provides a similar analysis to Labbe et al.\ for RUBIES-BLAGN-1 ($z=3.1$). We will present their spectral energy distributions (SEDs) in \S \ref{sec:sed} and bolometric corrections in \S \ref{sec:bolometric}. 

\subsection{MIRI, ALMA, and ancillary IR data}

In order to constrain the IR SED of these luminous LRDs, \cite{Setton:2025} compiled existing IR limits and obtained new ALMA and MIRI observations for these two luminous LRDs. A2744-45924 was observed in MIRI/F1000W and MIRI/F2100W for 11 and 30 minutes, respectively (JWST/GO \#6761, PI: Greene), resulting in detections that ruled out the presence of hot-dust from a torus \citep{Setton:2025}. Additionally, \cite{Setton:2025} present Herschel/PACS non-detections at 100 and 160 $\mu$m, based on imaging from the Herschel Lensing Survey \citep{Egami:2010}. ALMA Band 6 non-detections with 2 hour integrations are presented in \cite{fujimoto:2023dualz}. Finally, \cite{Setton:2025} present deep non-detections in Band 7 and Band 9 in 97 and 99 minute integrations, respectively.

\cite{Wang:2024brd} present MIRI/F770W and MIRI/F1800W detections of RUBIES-BLAGN-1 from the PRIMER survey (JWST/GO \#1837; PI: Dunlop), similarly finding that there is little evidence of a dominant torus. They also present Spitzer/MIPS 24 $\mu$m imaging \citep{Dickinson:2003} and Herschel/PACs 100 and 160 $\mu$m imaging from the 3D-Herschel project (S. McNulty et al. in preparation, NASAADAP-80NSSC20K0416), where the source is not detected \citep{Wang:2024brd, Barro:2023}. Finally, \cite{Setton:2025} present ALMA Band 6 and Band 8 non-detections based on 115 and 198 minute observations. The data are available at MAST: \dataset[doi: 10.17909/m7ks-wg55]{https://archive.stsci.edu/doi/resolve/resolve.html?doi=10.17909/m7ks-wg55}. 

These full IR SED constraints, in addition to NIRSpec/PRISM spectroscopy and X-ray non-detections, are shown in Figure \ref{fig:MonsterSED}. As we will argue more fully in \S \ref{sec:intrinsic}, the broad-band SEDs of A2744-45924 and RUBIES-BLAGN-1 are quite representative of the broader class of Little Red Dots, particularly spectroscopically selected objects. In terms of photometrically selected objects, \citet{Hviding:2025} present a detailed look at both the purity and completeness of a range of common color selections, as compared with a spectroscopic selection based on v-shape, compactness, and broad Balmer lines. They show that photometric selections tend to be relatively pure, but are incomplete in different ways. Thus, focusing on spectroscopic samples provides the most complete view of the Little Red Dots as a class. We also show the maximum IR luminosity allowed by the ALMA $3~\sigma$ upper limits in light purple, which is derived by assuming that the dust SED consists of a series of modified blackbodies.

\subsection{Gravitational lensing correction} \label{sec:SL}
To account for the gravitational magnification of the foreground cluster Abell\,2744, we use the \texttt{v2.0} UNCOVER lensing model, initially constructed by \citet{Furtak:2022b_SLmodel} and updated with new spectroscopic redshifts from JWST in \citet{Price:2024}. The magnifications are calculated at each object's position and redshift and then corrected for in the luminosity calculations.

We also use the lensing model to measure the volume used in our luminosity function in \S~\ref{sec:LF} by computing the cumulative source plane area as a function of magnification in each redshift bin. This area is then integrated with the differential volume element to compute the effective survey volume corrected for lensing.

\begin{table*}
\hskip -10mm
\begin{tabular}{l|l|c|ccc|c|c}
\hline
Region & $\lambda_{\rm rest}$ & $L_{\rm bol}/L_{\nu}$ & & $L_{\rm bol}/L_{\nu}$-minFIR & &  $L_{\rm bol}/L_{\nu}$-maxFIR & Ref. \\ 
  & $\micron$ &  Standard &  45924 & RUBIES-BLAGN-1 & Ave  & Ave &  \\
  (1) & (2) & (3) & (4) & (5) & (6) & (7) & (8) \\
\hline
$L_{X}$   & 0.0001--0.001 & 25 & $>$53 & $>$33 & $>$43  & $>$288 & RE04\\ 
$L_{\rm UV}$  & 0.00125--0.3 & 2  & 7    & 36    & 22    &  149 & RE04 \\
$L_{\rm opt}$ & 0.3--1    &  8 & 2.7  & 3.4   & 3.1   &  21 & RE04  \\
$L_{5100}$    & 0.51        &  9 & 3.9  & 6.9   & 5.4   &  37 & K00  \\
$L_{\rm H \alpha}$ & 0.65  & 170 & 11  &  30   & 19   & 131 & GH05 \\
$L_{\rm NIR}$     & 1--3 & 14 & 3.3   & 2.6   & 2.9   & 20 & RE04 \\
$L_{\rm IR}$  & 3--10$^5$ & 3.3 & 5.6 & 3.8 & 4.7  & $>1.2$ &  RE04 \\
\hline
\end{tabular} 
\caption{Luminosity in different wavelength ranges, and resulting bolometric corrections. The columns are:
(1) The wavelength region name, matched to Fig.\ \ref{fig:MonsterSED}.
(2) The wavelength range integrated over, adopted from the References in column (8).
(3) The standard bolometric correction in each wavelength region. 
(4) Bolometric corrections in each wavelength region derived for A2744-45924. 
(5) Bolometric corrections in each wavelength region derived for RUBIES-BLAGN-1.
(6) The average of (4) and (5); the default bolometric corrections adopted in this paper.
(7) Comparable to (6), the average bolometric correction across A2744-45924 and RUBIES-BLAGN-1, but adopting the maximum FIR in the bolometric luminosity.
(8) Reference for the ``standard'' bolometric correction. RE04: \citet{RisalitiElvis:2004}; K00: \citet{Kaspi:2000}; GH05: \citet{GreeneHo:2005}; L09: \citet{Liu:2009}. 
}

\label{tab:bolcor}
\end{table*}

\section{Spectral Energy Distributions}
\label{sec:sed}

In this section, we describe how we calculate bolometric luminosities for our two sources. While we have quite broad X-ray to far-infrared coverage for these two objects, we cannot directly observe the far-UV or most of the far-infrared emission. Without a complete model for the SED, we will adopt two possible extrapolations for the far-infrared. We will focus on a minimum model that assumes a small fraction of the total light emerges in the far-UV or far-infrared because one of our main goals is to explore the implications for Little Red Dot demographics if there is no hidden luminosity in the far-infrared. In that case, much lower bolometric corrections would imply lower black hole masses than have been previously published.

We begin with the most optically luminous known Little Red Dot, A2744-45924 \citep{Labbe:2025}. The SED measurements are summarized in Figure \ref{fig:MonsterSED}. \citet{Setton:2025} put conservative upper limits on the possible emission from the mid-infrared to the far-infrared for A2744-45924, concluding that at most $\sim 2 \times 10^{12} \, L_{\odot}$ could be emerging at long wavelengths. We have deep X-ray limits of $<2 \times 10^{43}$~erg\,s$^{-1}$ between 10--40~keV \citep{Labbe:2025}. Since this X-ray upper limit is five times fainter than the observed H$\alpha$ luminosity, we consider the X-ray contribution to $L_{\rm bol}$ to be negligible. Although we do not have radio constraints for this source, we also consider it likely that the radio contribution to $L_{\rm bol}$ is negligible, given the many non-detections for other sources in the literature  \citep{Akins:2024,Perger:2025,Gloudemans:2025}.

To calculate the bolometric luminosity, we take the observed PRISM spectrum, and extrapolate to X-ray wavelengths using the observed UV slope (purple line in Figure \ref{fig:MonsterSED}). The UV component thus calculated comprises a negligible fraction of $L_{\rm bol}$ (Table \ref{tab:bolcor}), and so we accept that the UV likely will include some galaxy contribution \citep[e.g.,][]{Chen:2024,Torralba:2025}. If there were a substantial UV bump at shorter wavelength that we do not see, then a good fraction of that bump should be absorbed and re-emitted in the mid-to-far--infrared, and thus is included implicitly in our upper limits.

We also extend into the mid-infrared by interpolating between the MIRI detections. We perform a linear extrapolation from the reddest MIRI band to zero at the wavelength of the most constraining ALMA band (band 7 for A2744-45924, band 6 for RUBIES-BLAGN-1), assuming negligible energy output at $\lambda_{\rm rest}>100$ $\mu$m (Figure \ref{fig:MonsterSED}). This is our minimum FIR SED, which is loosely motivated by models in which the AGN is enshrouded in a dense gas, leading to black body-like emission at $4000-6000$~K \citep[\S \ref{sec:models}; e.g.,][]{Liu:2025lrd,Begelman:2025}. Taking this linear extrapolation gives us a minimum luminosity. Integrating the full SED yields a bolometric luminosity of $L_{\rm bol} = 1.1 \times 10^{45}$~erg/s. This luminosity represents the total light emerging in the UV/optical part of the spectrum for A2744-45924, in the only region of the spectrum where Little Red Dots have been detected \citep[e.g.,][]{Akins:2024,Setton:2025}. We infer 25 times less bolometric luminosity than was published in \citet{Greene:2024}, using bolometric corrections for standard AGN \citep[][]{RisalitiElvis:2004} that assume an underlying SED as shown in red in Figure \ref{fig:MonsterSED} combined with dust reddening. 

We determine the default (minimal) bolometric luminosity by integrating from the far-UV to the far-infrared using the extrapolation of the MIR slope further to the far-infrared (see Table \ref{tab:bolcor}), which corresponds to a case where very little UV has been reprocessed to far-infrared. In this minimal case, the entire population is much less luminous than has been previously inferred. We note that for different assumptions about the nature of dust, a significant part of the luminosity may still emerge in the infrared \citep[still consistent with the FIR upper limits in Table \ref{tab:bolcor};][]{Chen:2025dust}. For completeness, we also present the case that there is significant far-infrared emission lurking just below our upper limits, represented by the light purple peak in Figure \ref{fig:MonsterSED} (see also \S \ref{sec:systematics}). Recently discovered low-redshift Little Red Dots do show evidence of some $\sim 300$~K gas \citep{Lin:2025lowz}, which we cannot rule out in our $z=3-4.5$ objects, but which is quite sub-dominant in terms of total luminosity. More detailed models are still needed to understand the origin of the MIR emission in the $z \sim 0.1$ Little Red Dot analogs. 

We repeat the above analysis for RUBIES-BLAGN-1 \citep{Wang:2024brd,Setton:2025}, using our comparably deep MIRI and ALMA constraints on the mid-to-far infrared SED. The largest SED difference between the two objects is that RUBIES-BLAGN-1 is redder in the UV/optical, but since the UV component of the SED contributes negligibly to the bolometric luminosity, overall the bolometric corrections are very similar. We average the two objects and tabulate them in Table \ref{tab:bolcor}.

\section{Bolometric Corrections}
\label{sec:bolometric}

A fundamental property of all accreting black holes is their bolometric luminosity. We have long known that AGN show a wide range of SED shapes \citep[e.g.,][]{Elvis:1994}, and that the SEDs correlate with the Eddington ratio \citep[e.g.,][]{ho2008,VasudevanFabian:2007,Kubota:2019,Richards:2006sed}. In general, we only have access to a small part of the spectrum, and so ``typical'' bolometric corrections are very commonly adopted. For instance, $L_{\rm bol} = 9 L_{5100}$, where $L_{5100}$ is defined as $\nu L_{\nu}$ at 5100\AA\ is often used for rest-frame optical spectra \citep[e.g.,][]{Kaspi:2000}. These corrections assume a standard AGN SED, with known $L_{\rm bol}$, and simply scale that value to the observed waveband. Such monochromatic values are preferred to broad-band magnitudes because they are line-free. 

Sometimes, integrated line emission is also used as a bolometric indicator. For instance, \citet{GreeneHo:2005} present a conversion from $L_{\rm H\alpha}$ to $L_{5100}$ for use in cases where the continuum luminosity from the AGN is ambiguous to measure---for instance because of host galaxy contributions---but the line is more accessible. This conversion is possible in standard AGN because the ratio of broad Balmer line flux to continuum, the EW, is nearly constant across objects \citep[e.g.,][]{Yee:1980,Shuder:1981,Stern:2012a}. As we discuss further in \S \ref{sec:table}, Little Red Dots have much higher EWs of 400--1000~\AA\ \citep[e.g.,][]{Lin:2024aspire} compared to values of 100--200~\AA\ for standard AGN \citep[e.g.,][]{VandenBerk:2001,Croom:2002,Stern:2012a}. Therefore, the bolometric corrections derived from H$\alpha$ are even more systematically offset for Little Red Dots relative to the standard AGN.

In this section, we revisit bolometric corrections for Little Red Dots, acknowledging that they very likely do not have a standard AGN SED.


\subsection{The assumption of an intrinsically red SED}
\label{sec:intrinsic}

In this work, we play out the hypothesis that the Little Red Dot SED is dominated by the light that we see in the UV/optical. We have a number of reasons to prefer a picture in which the red continuum that we see is intrinsic to the Little Red Dot, rather than reddened by dust. Extensive modeling efforts to describe the rest-frame UV/optical SEDs with a combination of standard AGN and galaxy templates were unable to find satisfactory solutions \citep[e.g.,][]{Wang:2024brd, Ma:2024lens, Wang:2024UB}. Most dramatically, \citet{deGraaff:2025} show that there are no known stars that can explain the sizable break measured in the RUBIES ``Cliff''. 

We also prefer an intrinsically red continuum due to the lack of detected hot or cold dust in two luminous Little Red Dots \citep{Setton:2025} and in stacks of larger numbers of less luminous sources \citep{Akins:2024,Casey:2025}. The lack of reprocessed emission strongly disfavors significant dust-reddened UV emission from either star formation or a standard AGN \citep{Setton:2025}. The fact that none of the targets have detections strongly argues that our SEDs are representative of the larger population. Finally, line ratios consistent with no dust reddening have also been seen in the narrow-line regions of some Little Red Dots \citep{Tang:2025,Lin:2025lowz}, suggesting little dust on larger scales. While some objects classified as Little Red Dots may be dusty, this paper focuses on the bulk of the population that cannot be explained as reddened by dust. In \S \ref{sec:models}, we discuss possible models to explain these SEDs.

\subsection{Table of bolometric corrections}
\label{sec:table}

We now calculate LRD-specific bolometric corrections for key wavelengths, again under the assumption that what we see in the UV/optical dominates the bolometric luminosity (Table \ref{tab:bolcor}). We do not apply any reddening corrections in calculating these bolometric corrections, as justified in \S \ref{sec:intrinsic}. We avoid UV corrections for now, both because the sources are very faint in the rest-frame UV and because the origin of the UV may well vary from object to object \citep[e.g.,][]{Torralba:2025,Chen:2025image}.
Examining each wavelength in turn, we see that the ratio of $L_{\rm bol} / L_{5100}$ is a factor of two lower in Little Red Dots than in standard AGN \citep[e.g.,][]{Richards:2006sed, VasudevanFabian:2007}. The ratio $L_{\rm bol} / L_{\rm NIR}$ is considerably lower, because so much of the total emission emerges in the rest-frame optical. Most dramatically, we find that the $L_{\rm bol} / L_{\rm H\alpha}$ luminosity is more than $10$ times lower than in standard AGN.

Beyond the differences in SED, there are two additional contributing factors causing the H$\alpha$ bolometric correction in particular to be so dramatically different from that of standard AGN \citep[e.g.,][]{GreeneHo:2005}. The first factor is that some works \citep[e.g.,][]{Furtak:2023nature,Kokorev:2023,Kokorev:2024,Kocevski:2024,Greene:2024} apply a steep dust correction with $A_V \sim 1$--3~mag, adding a factor of 2--5 overestimate of the bolometric luminosity. The second factor, alluded to at the beginning of this section, is that the \citet{GreeneHo:2005} relations implicitly assume a constant H$\alpha$ EW, which allows the H$\alpha$ line luminosity to be a proxy for $L_{5100}$. Contrary to typical AGN, the observed EWs of the Little Red Dots show H$\alpha$ EWs that are factors of 2--5 higher \citep[e.g.,][]{Setton:2024break,Lin:2024aspire}, artificially increasing the bolometric corrections for Little Red Dots compared to standard AGN. 

\subsection{Impact of new bolometric corrections}

\begin{figure*}
\vspace{5mm}
\includegraphics[width=0.5\textwidth]{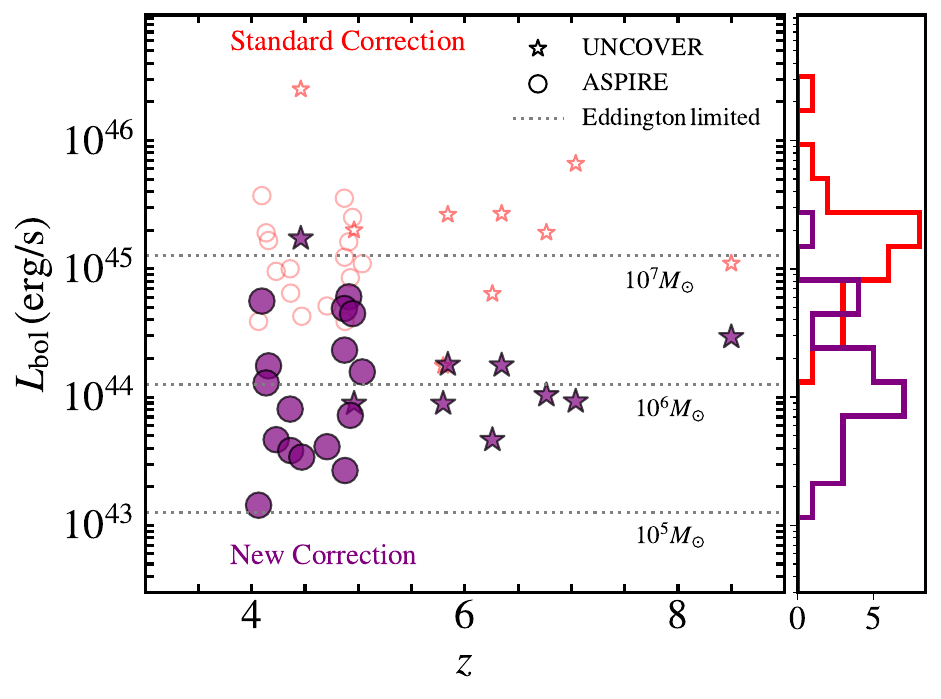}
\hspace{3mm}
\includegraphics[width=0.4\textwidth]{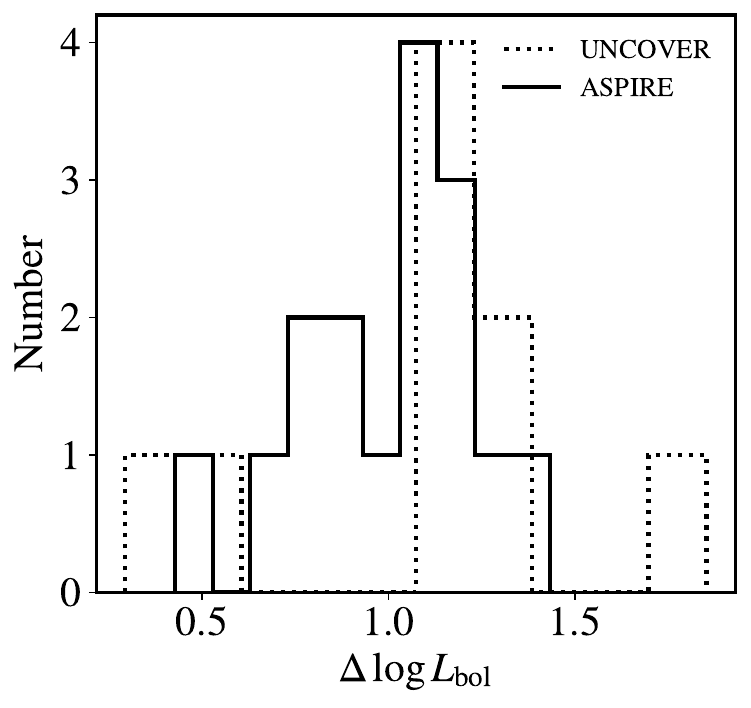}
\caption{{\it Left}: Published values of $L_{\rm bol}$ based on H$\alpha$ (red) compared with new values presented here (purple). We show measurements from ASPIRE \citep[][circles]{Lin:2024aspire} and UNCOVER \citep[][stars]{Labbe:2023uncover,Greene:2024}. In the case of ASPIRE, the H$\alpha$ luminosities were taken with no additional reddening correction, while in the case of UNCOVER, a typical $A_V \sim 1.5$~mag correction was applied. Across the board, we see that the implied bolometric corrections shift downwards by an order of magnitude for both samples. We also indicate the Eddington limit for black holes with \mbh$=10^5, \, 10^6, \, 10^7$~\msun. If we were to assume that all Little Red Dots were radiating at around their Eddington limit, then the implied black hole masses are $\sim 10^5$--$10^7$~\msun.
{\it Right}: The distribution in $\Delta$ log $L_{\rm bol}$, showing that the old correction including an extinction correction leads to heavily inflated bolometric luminosities.
}
\label{fig:zlbol}
\end{figure*}

The ramifications of the downward shift in bolometric luminosity are shown graphically in Figure \ref{fig:zlbol} as a function of redshift (left) and distribution of offsets (right). Two illustrative samples are chosen to demonstrate the dramatic difference in bolometric luminosity between the standard H$\alpha$-based calculation and the new lower values. We focus on one sample where a reddening correction was applied to the H$\alpha$ luminosities, the PRISM sample from UNCOVER \citep{Greene:2024}. For this sample, we adopt the ``no-FIR'' bolometric correction (Table \ref{tab:bolcor}) using $L_{5100}$ measured from the UNCOVER/PRISM spectra. We also include the ASPIRE grism sample \citep{Lin:2024aspire}. In the case of the ASPIRE objects, we only have reliable H$\alpha$ emission, so we again use the ``no-FIR'' bolometric correction based on H$\alpha$ (see \ref{sec:haewcor} for a minor correction). Even with no additional dust correction, the bolometric luminosity drops by an order of magnitude from the published values.  

We do not have direct information about the black hole masses of these objects. We do know that using standard scaling relations will not provide reliable broad-line region radii, given the very different H$\alpha$ EWs described above, and the non-standard relationship between the optical and bolometric luminosities. We do not know if the line widths are dominated by virial motions, turbulence, or scattering \citep[e.g.,][]{Rusakov:2025,Naidu:2025,Juodzbalis:2025}. However, if the bolometric luminosities are lower by more than an order of magnitude relative to published values, then very likely the black hole mass scale must also be dramatically lowered. To illustrate this idea, we highlight the Eddington luminosity for black holes with \mbh$=10^5, 10^6, 10^7$~\msun. Even without invoking super-Eddington accretion \citep[e.g.,][]{Lambrides:2024,Trinca:2024}, the typical masses for Little Red Dots seem likely to be \mbh$\sim 10^5$--$10^7$~\msun\ rather than \mbh$\sim 10^6$--$10^8$~\msun. Such black holes are hosted in $M^* \sim 10^8-10^{10}$~\msun\ galaxies locally \citep[e.g.,][]{reinesvolonteri2015,sagliaetal2016,Greene:2020}.

Next, we investigate systematic uncertainties in the bolometric corrections, both due to differences in SEDs and due to the range in H$\alpha$ EWs observed across the full Little Red Dot sample (\S \ref{sec:systematics}). 

\section{Systematic Uncertainties}
\label{sec:systematics}

We now address systematic uncertainties in the bolometric corrections. First, our default assumption is that there is no reprocessed emission in the far-infrared. This is likely too extreme, as there may well be some dust reprocessing even if the intrinsic emission is red and peaking in the rest-frame optical. We therefore present an upper limit to that reprocessed light (Table \ref{tab:bolcor}). Second, we consider heuristically the range of SEDs in the rest-frame optical for a large sample of Little Red Dots with rest-optical spectra \citep[][de Graaff in preparation]{Hviding:2025}, and how much variance is added to the bolometric corrections. Finally, we consider the range of H$\alpha$ EWs, and the additional uncertainty added in the $L_{\rm H\alpha}/L_{\rm bol}$ correction specifically.

\subsection{Far-infrared contributions}

We estimate the maximum amount of emission that we could be missing in the far-IR. \citet{Setton:2025} provide an empirical upper limit to the possible reprocessed emission by fitting a sum of black bodies that is consistent with all published upper limits in the mid-to-far infrared (light-purple line in Figure \ref{fig:MonsterSED}). Including that emission yields an upper limit of $L_{\rm bol} < 7 \times 10^{45}$~erg/s. Note that this estimate is completely dominated by the upper limits in the far-infrared, which have relatively low constraining power. We perform the same exercise for RUBIES-BLAGN-1, and the ``FIR'' correction in Table \ref{tab:bolcor} represents the maximum possible bolometric correction.

It is also worth noting that a larger number of Little Red Dots have ALMA observations covering a similar rest wavelength range (around $\sim 100 \, \micron$) as we have for RUBIES-BLAGN-1 and A2744-45924 \citep[e.g.,][]{Xiao:2025}. While none are as intrinsically luminous as our two sources, \citet{Casey:2025} perform a stacking analysis of their non-detections and infer a nearly identical ratio of rest-frame optical to far-infrared luminosities from the aggregate population as come from the two very luminous sources. While we do see a variety of Balmer break strengths and emission line properties across the Little Red Dot population \citep[e.g.,][]{Tang:2025,deGraaff:2025}, implying real variation in intrinsic SED, we believe that the far-IR SED limits inferred from the two most luminous targets presented here are fully consistent with existing constraints to the population as a whole.

\subsection{Range in optical SEDs}

Even if we accept that all of the emission for the Little Red Dots emerges in the rest-frame optical/UV, there is a variation in rest-frame optical spectral properties across a larger spectroscopic Little Red Dot sample \citep[e.g.,][]{Setton:2024break,Hviding:2025}. This range translates into a range of bolometric corrections based on $L_{5100}$. To estimate the magnitude of the range, we adopt modified black body fits to the rest-frame optical region (0.42-1$\micron$) from de Graaff et al.\ in preparation. These fits allow us to investigate how the ratio $L_{5100}/L_{\rm opt}$ varies with the apparent black body temperature. We find on average that $L_{5100}/L_{\rm opt} = 0.4 \pm 0.15$. Considering that the optically emitting part of the spectrum accounts for $\sim 30\%$ of the bolometric luminosity (Table \ref{tab:bolcor}), we estimate a factor of two spread in $L_{\rm bol}/L_{5100}$.  We note that our two sources span the distribution, with $L_{5100}/L_{\rm bol} \approx 0.3$ for A2744-45924 and $L_{5100}/L_{\rm bol} \approx 0.14$ for RUBIES-BLAGN-1.  Future work with larger samples can correct this small systematic uncertainty, while De Graaff et al.\ in preparation will investigate in detail how optical color correlates with the luminosity and line properties of the full spectroscopic Little Red Dot sample.

\subsection{Range in H$\alpha$ Equivalent Widths}
\label{sec:haewcor}

In addition to the range in observed rest-optical SED shapes, Little Red Dots demonstrate a wide range (factor of four) in rest-frame H$\alpha$ EW compared to standard AGN \citep[e.g.,][]{VandenBerk:2001,GreeneHo:2005,Stern:2012a,Lin:2024aspire}. The very constant EW of standard AGN is taken as evidence that photoionization dominates the excitation of typical broad lines \citep[e.g.,][]{Searle:1968}. Little Red Dots have EWs that are considerably higher but also show a much wider range \citep[e.g.,][]{Matthee:2023,Lin:2024aspire}, which is an unexplained puzzle. From a practical perspective, this wide range of EW makes H$\alpha$ a poor bolometric indicator in Little Red Dots. To mitigate this variance, one can normalize the EW of their target Little Red Dot by 940\AA, which is the average H$\alpha$ EW of RUBIES-BLAGN-1 and A2744-45924. This renormalization will ensure that on average H$\alpha$ and $L_{5100}$ will give the same $L_{\rm bol}$.


\section{Demographic Implications}
\label{sec:demographics}

In this section, we investigate the number densities and mass scale of Little Red Dots in light of the new bolometric luminosities. Bolometric luminosities are key to establishing the mass scale of AGN. Since BHs only occupy some range of Eddington ratio, the bolometric luminosities imply the distribution of black hole mass, stellar mass, and halo mass \citep[e.g.,][]{Volonteri:2017}. 

\begin{figure*}
\vspace{-1mm}
\includegraphics[width=0.40\textwidth]{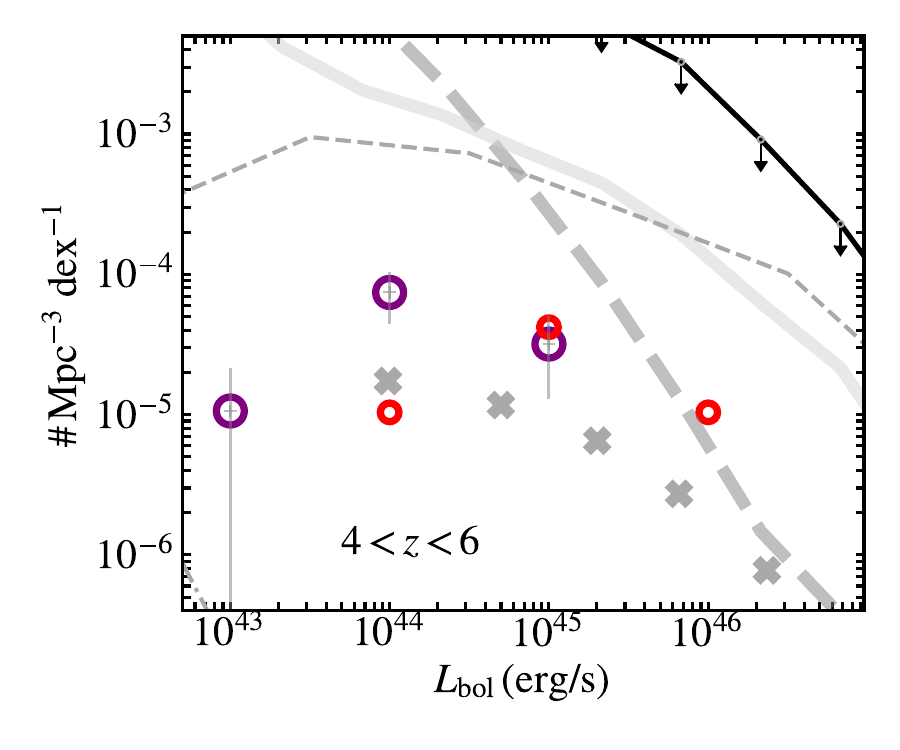}
\includegraphics[width=0.625\textwidth]{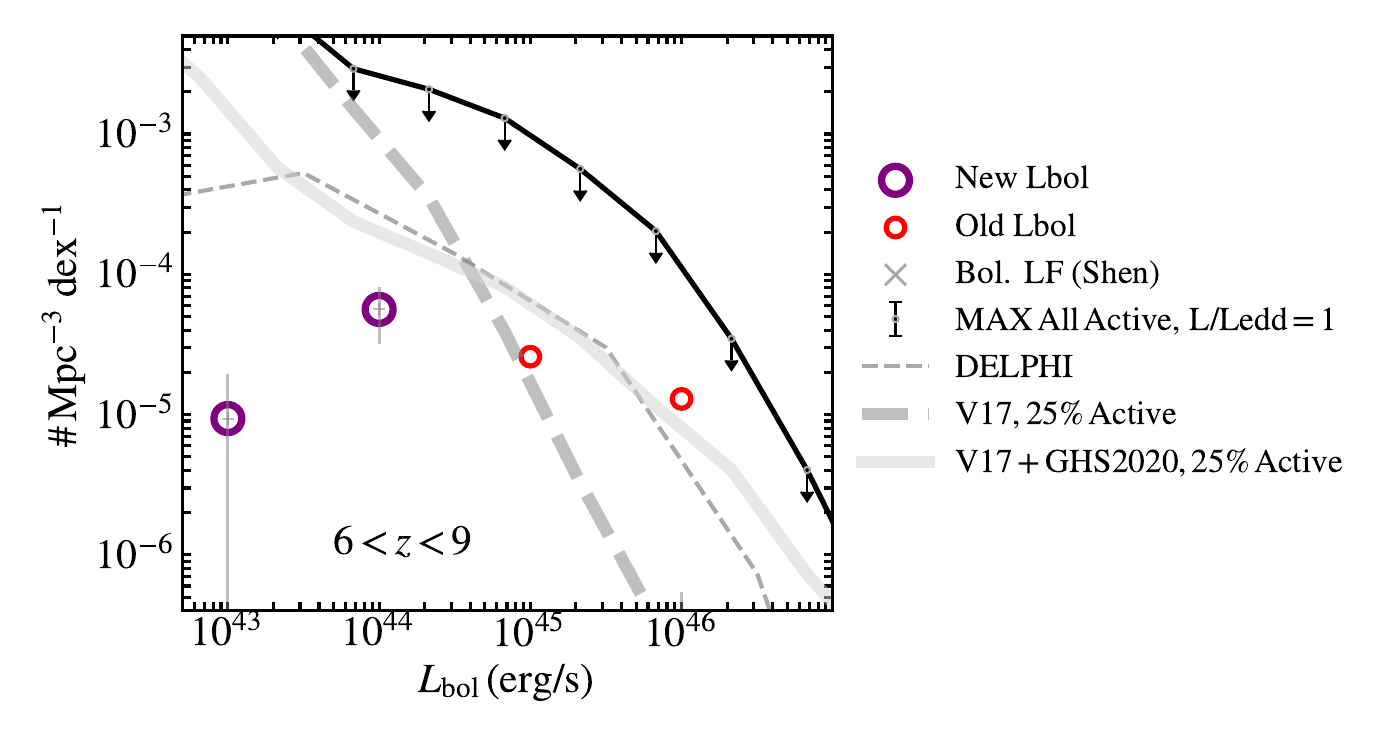}
\caption{Bolometric luminosity function from the current analysis (purple symbols) as compared with \citet{Greene:2024} (red). For reference, we show the ``maximum'' line, which assumes that every halo hosts a black hole radiating at its Eddington limit. The points are still higher than the bolometric luminosity function based on X-ray and UV observations from \citet[][grey crosses]{Shen:2020}, but now sit comfortably below both the DELPHI models \citep{Dayal:2025reion}, and models from \citet{Volonteri:2017} assuming two different scaling relations, derived from AGN \citep{reinesvolonteri2015} and from dynamical BH masses \citep[GHS][]{Greene:2020}, with a 25\% active fraction. At $z>6$, the \citet{Greene:2024} values exceed the model lines and approached the maximum line. That tension is alleviated with the new $L_{\rm bol}$ values. }
\label{fig:LF}
\end{figure*}

\subsection{Bolometric Luminosity Functions} \label{sec:LF}

One of the starkest puzzles about the Little Red Dots has been their high number density at relatively high inferred bolometric luminosity. Prior to \emph{JWST}, there was an apparent dearth of accreting black holes at $z>5$ compared to expectations, as demonstrated by the \citet{Shen:2020} luminosity function in Figure \ref{fig:LF}  \citep[see][ for a wider family of models]{Amarantidis2019,habouzitLF:2022}. The number densities of Little Red Dots imply high occupation fractions and duty cycles at elevated Eddington ratios \citep[e.g.,][]{Greene:2024}. Little Red Dots comprise only a fraction of broad-line objects \citep[][]{Hviding:2025}. However, \citet{Greene:2024} find that luminous Little Red Dots at $6<z<9$ are only ten times less common than a scenario in which every halo hosts an Eddington-limited accreting black hole (see the ``maximum'' line in Figure \ref{fig:LF}). Such a high number density at relatively high $L_{\rm bol}$ is uncomfortable. Thus, we revisit the bolometric luminosity function \citep{Greene:2024,Kokorev:2024} to evaluate the extent to which this tension is alleviated.

In Figure \ref{fig:LF}, we present a revised bolometric luminosity function in the Abell~2744 field, based on UNCOVER \citep{Bezanson:2022} photometry \citep{Labbe:2023uncover,Greene:2024} and spectroscopy \citep{Price:2024}, as well as ALT broad-line selection \citep{Matthee:2024}. Our definition of Little Red Dot follows that of \citet{Hviding:2025}: a target should have a v-shaped continuum, a dominant point-source component in the rest-frame optical, and broad Balmer lines to be a spectroscopically identified Little Red Dot. The fraction of broad-line selected objects that qualify as Little Red Dots varies from 20--70\% depending on whether objects are continuum or H$\alpha$ selected and on the line sensitivity \citep[e.g.,][]{Harikane:2023,Matthee:2023,Lin:2024aspire,Hviding:2025}.

To build the bolometric luminosity functions, we start with the $f_{5100}$ values. We derive the new bolometric luminosities from the 5100~\AA\ fluxes, as measured from the UNCOVER spectra directly. We add three new sources from the All the Little Things \citep{Naidu24} grism spectroscopy, as presented in \citet{Matthee:2024env}. Since we do not detect the continuum in the grism spectroscopy, we are obliged to use medium-band imaging from the Cycle 2 program Medium-bands, MegaScience \citep[JWST-GO-4111, PI: K.\ Suess;][]{Suess:2024}. For this purpose, we simply select line-free medium bands straddling rest-frame 5100\AA. We fit a power-law to the flux densities and then derive the 5100\AA\ value from this simple continuum fit. In the three cases with overlap between UNCOVER-PRISM spectroscopy and ALT sources, we find agreement within a factor of two or better in all cases.

We then recalculate the bolometric luminosity functions presented in \citet{Greene:2024}, using the same completeness correction approach as described in that work. The volume is computed as described in section~\ref{sec:SL}, to which we add a 14\,\% uncertainty in order to account for lensing systematics (see Chemerynska et al. in prep.). The luminosity function uncertainties are then obtained by allowing Little Red Dots to change luminosity bin according to their bolometric luminosity uncertainties and combining this with the volume error and Poissonian noise. We present the revised bolometric luminosity functions in Figure \ref{fig:LF}. \citet{Hviding:2025} find that the color selection of \citet{Labbe:2023uncover} is roughly 50\% complete, with the UV-faintest sources systematically excluded; this incompleteness is not accounted for. 

At all redshifts, the typical object drops in luminosity such that the luminosity functions basically shift to lower luminosity by roughly a factor of ten (Figure \ref{fig:zlbol}). The visual impression is a bit messier because of slight differences in binning, but we see that the typical $z \sim 5$ object now has $L_{\rm bol} \approx 10^{45}$~erg/s and a number density of $n \approx 7 \times 10^{-5}$~Mpc$^{-3}$~dex$^{-1}$. At $6<z<9$, the typical object has $L_{\rm bol} \approx 10^{44}$~erg/s, and $n \approx 6 \times 10^{-5}$~Mpc$^{-3}$~dex$^{-1}$. 

The revised bolometric corrections completely alleviate the apparent high duty cycle. Now, our points fall below the predictions of \citet{Volonteri:2017}, which assume 25\% active fraction, both using the \citet{reinesvolonteri2015} and \citet[][GHS]{Greene:2020} local scaling relations (dashed and solid respectively). Likewise, the objects fall well below the predictions of the semi-analytic model DELPHI with \jw\ constraints folded in \citep{Dayal:2025reion}. We no longer appear to be overproducing black hole mass density \citep[e.g.,][]{InayoshiIchikawa:2024}, but we potentially return to the challenge that there are not enough known accreting black holes compared to standard predictions \citep[e.g.,][]{Volonteri:2017} as illustrated by the thick solid and dashed gray lines in Figure \ref{fig:LF}. We estimate that including all known broad-line selected sources would only change the number densities by $\sim$ a factor of 2--3 \citep[e.g.,][]{Harikane:2023,Hviding:2025}. Perhaps the solution will come from robustly incorporating truly obscured sources \citep[e.g.,][]{Scholtz:2024,Treiber:2024}, or perhaps our ``minimal'' luminosity model is too extreme.

\subsection{Lowering the implied mass scale}

We now explore the demographic implications of an order-of-magnitude downward shift in bolometric luminosity (Figure \ref{fig:zlbol}). We suggest that the black hole masses are likely to be shifted lower as well. We emphasize that we do not have reliable measurements of black hole mass for these objects, nor a reliable way to understand the size, density, or structure of the Balmer-line emitting region from which we measure velocity $v$. However, lowering the implied $L$ naturally lowers \mbh. Since black holes radiate at some range of luminosity given by the Eddington ratio distribution, lower $L_{\rm bol}$ naturally lowers the distribution of \mbh. Others have also suggested that the BH mass scale must be lower, either because super-Eddington accretion is likely at play \citep[e.g.,][]{lupi2024,Lambrides:2024}, or because the observed broad-line velocities may be broadened by scattering \citep{Killi:2024,Rusakov:2025,Naidu:2025}.

Positing lower \mbh\ would remove the significant tension in the black hole to galaxy ratios \citep[e.g.,][]{Furtak:2023nature, Kokorev:2023, Maiolino:2023}. While we do not have robust stellar masses for these sources, we do have reasonable estimates from their small sizes and low apparent dynamical host galaxy masses based on \oiii\ \citep{Wang:2024UB, Ji:2025}. Compact morphologies and low UV luminosities also argue for low-mass hosts \citep[e.g.,][]{Chen:2025image,Torralba:2025,Naidu:2025}. Furthermore, clustering results \citep{Pizzati:2025,Matthee:2024,Lin:2025} and observed number densities both suggest that Little Red Dots generally cluster like relatively low-mass galaxies (e.g., M$_* \sim 10^7$--$10^8$~\msun). One possible exception is A2744-45924 itself, which is in an overdensity at $z=4.46$ \citep{Labbe:2025}, but may not be the most massive galaxy \citep{Torralba:2025}. The high observed number densities are also much easier to accommodate when $L_{\rm bol}$ is lower by an order of magnitude, since as \mbh\ and $M_*$ get lower, we can expect these sources to occupy lower-mass and far more numerous dark matter halos (Fig. \ref{fig:LF}).  

Lowering the bolometric luminosities is also more comfortable from the perspective of the amount of mass density built up at $z>5$. \citet{InayoshiIchikawa:2024} find that the radiative efficiency in the Little Red Dots needs to be quite high or their early growth will overproduce the local black hole mass density. If the bolometric luminosities are overestimated by a large factor, then the implied mass density is no longer extreme \citep{Chen:2025dust}. 

\subsection{Models with Intrinsically Red SEDs}
\label{sec:models}

A productive way to think about the observed SEDs has come from thinking about the analogy between accretion disks and photospheres of stars \citep[e.g.,][]{Hubeny:2000,begelman2008,Begelman:2025}. There are theoretical models that successfully reproduce the observed spectrum by inserting a slab of high column-density gas between us and the accretion disk \citep{Inayoshi:2024,Ji:2025}, although in these models the incident spectrum is not calculated self-consistently, but may need to be intrinsically red to explain the observed SED without invoking significant reddening and non-standard dust laws \citep{Naidu:2025,deGraaff:2025}. 

The models presented by \citet{Liu:2025lrd} show that a quasi-spherical accretion flow accreting at super-Eddington rates can emit roughly as a black-body at $T_{\rm eff} \approx 5000$~K for a wide range of black hole masses and gas densities. In these models, rather than dust reprocessing, which would deposit emission into the infrared, the gas absorption we imagine here simply changes the dominant temperature of the emission at the photosphere. The model naturally reproduces the observed Balmer break similar to A stars, but at lower photosphere density and thus temperature \citep[see also quasi-star models, e.g.][]{begelman2008, Begelman:2025}. In light of these developments, it is valuable to consider a situation in which the bulk of the emission emerges in the UV/optical component. 

In particular, in the case of the Little Red Dots, a good match to the Balmer break and the peak wavelength in the rest-frame optical comes from an accretion flow that has a photosphere at $\sim 5000$~K  \citep[][dubbed a BH* by Naidu et al]{Inayoshi:2024, Ji:2025, Naidu:2025, deGraaff:2025, Kido:2025, Taylor:2025, Liu:2025lrd}. 

We can contrast our empirically constrained bolometric corrections with an example BH star (BH*) model selected to reproduce the rest-optical features of MoM-BH*-1 at $z=7.76$ \citep{Naidu:2025}, using the best-fit model presented in that paper. While the model is constructed to fit H$\beta$ in emission, H$\gamma$ in absorption, and the Balmer break (among the strongest ever observed along with the Cliff; \citealt{deGraaff:2025}), from the model we have predictions for the emission at longer (far-IR) and shorter (UV, X-ray) wavelengths. The total bolometric luminosity based on the model is $\log{(L_{\rm{bol}}/\rm{erg\ s^{-1}})}=44.5$, with $L_{\rm{bol}}/L_{\rm{opt}}=2.6$. This ratio is in agreement with the estimate in Table \ref{tab:bolcor}---i.e., the bulk of the total luminosity emerges in the rest-optical. The implied $L_{\rm bol}/L_{\rm H\alpha} = 39$ is also in line with the values presented here, whereas following the ``standard" approach commonly adopted in the literature, one would overestimate the luminosity by more than an order of magnitude $\log{(L_{\rm{bol}}/\rm{erg\ s^{-1}})}=45.7^{+0.1}_{-0.1}$ (via the dust-corrected H$\beta$ emission line, e.g., \citealt[][]{Vestergaard:2006}).

\section{Summary}

We present a very simple argument: if the Little Red Dots are not heavily dust reddened, but have intrinsically red rest-frame optical spectra, then published bolometric luminosities assuming a dust-reddened standard AGN or starburst are ten times too high. Accretion models of quasi-spherical flows predict SEDs similar to those that we observe \citep[e.g.,][]{begelman2008,Liu:2025lrd}, and also predict much lower bolometric luminosity \citep{Naidu:2025}.

The implied luminosity and mass scale for these objects shifts downward dramatically. In this case, which we believe is likely for a large fraction of the Little Red Dot population, the accreting black holes may well be found in plentiful, low-mass halos with host galaxies of $10^7-10^8$~\msun, and would represent an important early phase in black hole growth. Given that Little Red Dots are observed at least to $z \sim 9$ \citep{Taylor:2025}, they may become a critical tool in studying the progenitors of black hole seeds \citep[e.g.,][]{volonteri2010}.

\section*{Acknowledgements}

We benefit from the following JWST programs: UNCOVER (JWST/GO \#2561; Labb{\'e} \& Bezanson); ALT (JWST-GO \#3516; Naidu \& Matthee); MegaScience (JWST-GO \#4111; Suess); RUBIES ((JWST-GO \#4233; de Graaff \& Brammer); PRIMER (JWST/GO \#1837; Dunlop).

We acknowledge funding from NSF/AAG \#2306950, JWST-GO-02561, JWST-GO-03516, and JWST-GO-04111, provided through a grant from the STScI under NASA contract NAS5-03127. I.L. acknowledges support from Australian Research Council Future Fellowship FT220100798. K.G. and T.N. acknowledge support from Australian Research Council Laureate Fellowship FL180100060. AZ acknowledges support by Grant No. 2020750 from the United States-Israel Binational Science Foundation (BSF) and Grant No. 2109066 from the United States National Science Foundation (NSF); by the Ministry of Science \& Technology, Israel; and by the Israel Science Foundation Grant No. 864/23. JM and IK are funded by the European Union (ERC, AGENTS, 101076224). Views and opinions expressed are however those of the author(s) only and do not necessarily reflect those of the European Union or the European Research Council. Neither the European Union nor the granting authority can be held responsible for them.  YF ackowledges supports from JSPS KAKENHI Grant Number JSPS KAKENHI Grant Numbers JP22K21349 and JP23K13149.
This work has received funding from the Swiss State Secretariat for Education, Research and Innovation (SERI) under contract number MB22.00072, as well as from the Swiss National Science Foundation (SNSF) through project grant 200020\_207349.
The Cosmic Dawn Center (DAWN) is funded by the Danish National Research Foundation under grant DNRF140. Support for this work for RPN was provided by NASA through the NASA Hubble Fellowship grant HST-HF2-51515.001-A awarded by the Space Telescope Science Institute, which is operated by the Association of Universities for Research in Astronomy, Incorporated, under NASA contract NAS5-26555. The work of CCW is supported by NOIRLab, which is managed by the Association of Universities for Research in Astronomy (AURA) under a cooperative agreement with the National Science Foundation.  
JM acknowledges funding by the European Union (ERC, AGENTS,  101076224).
REH acknowledges support by the German Aerospace Center (DLR) and the Federal Ministry for Economic Affairs and Energy (BMWi) through program 50OR2403 ‘RUBIES’.

\bibliographystyle{aasjournal}
\bibliography{imbh.bib}

\end{document}